\documentclass{article}
\usepackage{amsmath}
\usepackage{amssymb}
\input epsf
\newcommand{\si}[1]{\sigma_{#1}}
\newcommand{\sa}[2]{\sigma_{#1}^{#2}}
\newcommand{\ip}[2]{\langle \,{#1},\,{#2}\,\rangle}
\newcommand{\ro}{\rho}
\newcommand{\la}{\lambda}
\newcommand{\al}{\alpha}
\newcommand{\be}{\beta}
\newcommand{\g}{\gamma_{0}}
\newcommand{\ga}{\gamma}

\newcommand{\re}{\mathrm{Re}\,}
\newcommand{\I}{\mathbb I}
\newcommand{\ket}[1]{|{#1}\rangle}
\newcommand{\bra}[1]{\langle {#1} |}
\newcommand{\cH}{{\mathcal H}}
\newcommand{\C}{\mathbb C}

\newcommand{\fA}{\mathfrak A}
\newcommand{\fE}{\mathcal E}
\newcommand{\tr}{\mathrm{tr}\,}
\newcommand{\ptr}[1]{\mathrm{tr}_{#1}}
\newcommand{\DS}{\displaystyle}

\begin{document}
\begin{center}
\begin{LARGE}
\textbf{Entangling  two-level atoms by spontaneous emission}
\end{LARGE}\\[12mm]
L. Jak{\'o}bczyk\\[2mm]
Institute of Theoretical Physics\\ University of
Wroc{\l}aw\\
Pl. M. Borna 9, 50-204 Wroc{\l}aw, Poland
\end{center}
\vskip 12mm
\noindent
{\sc Abstract}: It is shown that the dissipation due to
spontaneous emission can entangle two closely separated two-level
atoms.
\section{Introduction}
Analysis of various aspects of spontaneous emission by a system 
of
two-level atoms, initiated in the classical paper of Dicke
\cite{Dicke} was further developed by several authors (see e.g.
\cite{dill, lemb1, lemb2}). In particular, in the case of
spontaneous emission by two trapped atoms separated by a 
distance
small compared to the radiation wavelength,  where is a
substantial probability that a photon emitted by one atom will be
absorbed by the other, there are states of the system in which
photon exchange can enhance or diminish spontaneous decay 
rates.
The states with enhanced decay rate are called superradiant and
analogously states with diminished decay rate are called
subradiant \cite{Dicke}. It was also shown by Dicke, that the
system of two coupled two-level atoms can be treated as a single
four-level system with modified decay rates. Note also that such
model can be realized in a laboratory by two laser-cooled trapped
ions, where the observation of superradiance  and subradiance  is
possible \cite{DeVoe}.
\par
Another aspects of the model of the spontaneous emission
are studied in the present paper.
When the compound system of two atoms is in an entangled state,
the irreversible process of radiative decay usually destroys
correlations and the state becomes unentangled. In the model
studied here, the photon exchange  produces correlations between
atoms which can partially overcome  decoherence caused by
spontaneous radiation. As a result, some amount of entanglement
can survive, and moreover there is a possibility that this
process can entangle separable states of two atoms. The idea that
 dissipation can create entanglement in physical
systems, was recently developed in several papers
\cite{plenio,kim,milb,kni}. In the present paper we show that the
dissipation due to spontaneous emission can entangle two atoms
that are initially prepared in a separable state. We study the
dynamics of this process. In the Markovian approximation it is
given by the semi-group  $\{ T_{t} \}$ of completely positive
linear mappings acting on density matrices \cite{Alicki}. We
consider time evolution of initial state of the system as well as
the evolution of its entanglement, measured by so called
concurrence \cite{HW, W}, in the case when the photon exchange
rate $\ga$ is close to spontaneous emission rate of the single
atom  $\g$ and we can use the approximation $\g=\ga$ (similar
model was also considered in \cite{bash}). We calculate
asymptotic stationary states $\ro_{\mathrm{as}}$ for the
semi-group $\{ T_{t} \}$ and show that they depend on initial
conditions (i.e. $\{ T_{t} \}$ is relaxing but not uniquely
relaxing). The   concurrence of $\ro_{\mathrm{as}}$ also depends
on initial state and can be non zero for some of them. We discuss
in details some classes of initial states. In particular, we show
that there are pure separable states evolving to entangled mixed
states and such which remain separable during evolution. The first
class contains physically interesting initial state when one atom
in in excited state and the other is in ground state. The
relaxation process given by the semi-group $\{ T_{t} \}$ produces
in this case the states with entanglement monotonically
increasing in time to the maximal value. The class of pure
maximally entangled initial states is also discussed. Similar
"production" of entanglement is shown to be present for some
classes of mixed states. On the other hand, when the photon
exchange rate $\ga$ is
smaller then $\g$, the relaxation process brings all initial
states to the unique asymptotic state when both atoms are in its
ground states. As we show, even in that case the dynamics can 
entangle two
separable states, but the amount of entanglement is decreasing to
zero.
\section{Pair of two-level atoms}
Consider two-level atom $A$ with ground state $\ket{0}$
and excited state $\ket{1}$. This quantum system can be described
in terms of the Hilbert space $\cH_{A}=\C^{2}$ and the algebra
$\fA_{A}$ of $2\times 2$ complex matrices. If we identify
$\ket{1}$ and $\ket{0}$ with vectors $\bigl( \begin{smallmatrix}
1\\0
\end{smallmatrix}\bigr)$ and $\bigl( \begin{smallmatrix} 0 \\ 1
\end{smallmatrix}\bigr)$ respectively, then the raising and
lowering operators $\si{+},\;\si{-}$ defined by \begin{equation}
\si{+}=\ket{1}\bra{0},\quad \si{-}=\ket{0}\bra{1} \end{equation} can be
expressed in terms of Pauli matrices $\si{1},\; \si{2}$ 
\begin{equation}
\si{+}=\frac{1}{2}\,(\si{1}+i\,\si{2}),\quad
\si{-}=\frac{1}{2}\,(\si{1}-i\,\si{2}) \end{equation} For a joint system 
$AB$
of two two-level atoms $A$ and $B$, the algebra $\fA_{AB}$ is
equal to $4\times 4$ complex matrices and the Hilbert space
$\cH_{AB}=\cH_{A}\otimes \cH_{B}=\C^{4}$. Let $\fE_{AB}$ be the
set of all states of the compound system i.e. \begin{equation} 
\fE_{AB}=\{
\ro\in \fA_{AB}\, : \, \ro\geq 0\quad\text{and}\quad\tr \ro =1 \}
\end{equation} The state $\ro\in \fE_{AB}$ is \textit{separable}
\cite{Werner}, if it has the form \begin{equation}
\ro=\sum\limits_{k}\la_{k}\ro_{k}^{A}\otimes \ro_{k}^{B},\quad
\ro_{k}^{A}\in \fE_{A},\;\ro_{k}^{B}\in \fE_{B},\; \la_{k}\geq
0\quad\text{and}\quad \sum\limits_{k}\la_{k}=1 \end{equation} The 
set
$\fE_{AB}^{\,\rm sep}$ of all separable states forms a convex
subset of $\fE_{AB}$. When $\ro$ is not separable, it is called
\textit{inseparable} or \textit{entangled}. Thus \begin{equation}
\fE_{AB}^{\,\rm ent}=\fE_{AB}\setminus \fE_{AB}^{\,\rm sep} 
\end{equation}
If $P\in \fE_{AB}$ is a pure state i.e. $P$ is one-dimensional
projector, then $P$ is separable iff partial traces $\ptr{A}P$ and
$\ptr{B}P$ are also projectors. For mixed states, the separability
problem is much more involved. Fortunately, in the case of $4$ --
level compound system there is a simple necessary and sufficient
condition for separability: $\ro$ is separable iff its partial
transposition $\ro^{T_{A}}$ is also a state \cite{HHH}. Another
interesting question is  how to measure the amount of entanglement
a given quantum state contains. For a pure state $P$, the entropy
of entanglement \begin{equation} E(P)=-\tr[ (\ptr{A}P)\,\log_{2}\, 
(\ptr{A}P)]
\end{equation} is essentailly a unique measure of entanglement
\cite{poproh}. For mixed state $\ro$ it seems that the basic
measure of entanglement is the entanglement of formation
\cite{Bennett} \begin{equation} E(\ro)=\min \, 
\sum\limits_{k}\la_{k}E(P_{k})
\end{equation} where the minimum is taken over all possible 
decompositions
\begin{equation} \ro=\sum\limits_{k}\la_{k}P_{k} \end{equation} 
Again, in the case of
$4$ -- level system, $E(\ro)$ can be explicitely computed and it
turns out that $E(\ro)$  is the function of another useful
quantity $C(\ro)$ called \textit{concurrence}, which also can be
taken as a measure of entanglement \cite{HW, W}. Since in the
paper we use concurrence to quantify entanglement, now we 
discuss
its definition. Let \begin{equation} \ro^{\dag}=(\si{2}\otimes
\si{2})\,\overline{\ro}\,(\si{2}\otimes \si{2}) \end{equation} where
$\overline{\ro}$ is the complex conjugation of the matrix $\ro$.
Define also \begin{equation}
\widehat{\ro}=(\ro^{1/2}\ro^{\dag}\ro^{1/2})^{1/2} \end{equation} Then 
the
concurrence $C(\ro)$ is given by \cite{HW,W} \begin{equation} 
C(\ro)=\max\;
(\,0, 2p_{\mathrm{max}}(\widehat{\ro})-\tr \widehat{\ro}\,) 
\end{equation}
where $p_{\mathrm{max}}(\widehat{\ro})$ denotes the maximal
eigenvalue of $\widehat{\ro}$. The value of the number $C(\ro)$
varies from $0$ for separable states, to $1$ for maximally
entangled pure states.

\section{Decay in a system of  closely separated atoms}
We study the spontaneous emission of two atoms separated by a
distance $R$ small compared to the radiation wavelength . At such
distances there is a substantial probability that the photon
emitted by one atom will be absorbed by the other. Thus the
dynamics of the system is given by the  master equation
\cite{Agar} \begin{equation} \frac{d\ro}{dt}=L\ro,\quad \ro\in \fE_{AB} 
\end{equation}
with the following generator $L$ \begin{equation}
\begin{split}
L\ro=&\frac{\gamma_{0}}{2}\;[2\sa{-}{A}\;\ro\;\sa{+}{A}+2\sa{-
}{B}\;\ro\;\sa{+}{B}-
(\sa{+}{A}\;\sa{-}{A}+\sa{+}{B}\;\sa{-}{B})\;\ro-\ro\;
(\sa{+}{A}\;\sa{-}{A}+\sa{+}{B}\;\sa{-}{B})]+\\[2mm]
&\frac{\gamma}{2}\;[2\sa{-}{A}\;\ro\;\sa{+}{B}+2\sa{-
}{B}\;\ro\;\sa{+}{A}-(\sa{+}{A}\;
\sa{-}{B}
+\sa{+}{B}\;\sa{-}{A})\;\ro-
\ro \;(\sa{+}{A}\;\sa{-}{B}+\sa{+}{B}\;\sa{-}{A})]
\end{split}
\end{equation}
where
\begin{equation} \sa{\pm}{A}=\si{\pm}\otimes\I,\; \sa{\pm}{
B}=\I\otimes \si{\pm},\; \si{\pm}=\frac{1}{2}(\si{1}\pm i \si{2})
\end{equation}
Here $\gamma_{0}$ is the single atom spontaneous emission rate,
and $\gamma=g\gamma_{0}$  is a relaxation constant of photon
exchange. In the model, $g$ is the function of the  distance $R$
between atoms and $g\to 1$ when $R\to 0$. In this section we
investigate the time evolution of the initial density matrix
$\ro$ of the compound system, governed by the semi - group
$\{T_{t}\}_{t\geq 0}$ generated by $L$. In particular, we will
study the time development of entanglement of $\ro$, measured by
concurrence.
\par
Assume that the distance between atoms is so small that the
exchange rate $\gamma$ is close to $\gamma_{0}$ and we can 
use the
approximation $g=1$. Under this condition we study evolution of
the system and in particular we consider asymptotic states. Direct
calculations show that the semi - group $\{ T_{t} \}$ generated
by $L$ with $g=1$ is relaxing but not uniquely relaxing i.e.
there are as many stationary states as there are initial
conditions. More precisely, for a given initial state $\ro
=(\ro_{jk})$, the state $\ro(t)$ at time $t$ has the following
matrix elements
\begin{equation*}
\begin{split}
&\ro_{11}(t)=e^{-2\g t}\ro_{11}\\
&\ro_{12}(t)=\frac{1}{2}[e^{-2\g t}(\ro_{12}+\ro_{13})+e^{-
\g t}(\ro_{12}-\ro_{13})]\\
&\ro_{13}(t)=\frac{1}{2}[e^{-2\g t}(\ro_{12}+\ro_{13})+e^{-\g
t}(\ro_{13}-\ro_{12})]\\
&\ro_{14}(t)=e^{-\g t}\ro_{14}\\
&\ro_{22}(t)=\frac{1}{4}e^{-2\g t}(\ro_{22}+\ro_{33}+2\re
\ro_{23})+\frac{1}{2}e^{-\g t}(\ro_{22}-\ro_{33})+\g t  e^{-2\g
 t}\ro_{11}+\frac{1}{4}(\ro_{22}+\ro_{33}-2\re \ro_{23})\\
&\ro_{23}(t)=\frac{1}{4}e^{-2 \g t}(\ro_{22}+\ro_{33}+2\re
\ro_{23})+\frac{1}{2}e^{-\g t}(\ro_{23}-\ro_{32})+\g te^{-2\g
 t}\ro_{11}-\frac{1}{4}(\ro_{22}+\ro_{33}-2\re \ro_{23})\\
&\ro_{24}(t)=-e^{-2\g
 t}(\ro_{12}+\ro_{13})+\frac{1}{2}e^{-\g
 t}(2\ro_{12}+2\ro_{13}+\ro_{24}+\ro_{34})+\frac{1}{2}(\ro_{24}-
\ro_{34})\\
&\ro_{33}(t)=\frac{1}{4}e^{-2\g t}(\ro_{22}+\ro_{33}+2\re
\ro_{23})-\frac{1}{2}e^{-\g t}(\ro_{22}-\ro_{33})+\g t  e^{-2\g
 t}\ro_{11}+\frac{1}{4}(\ro_{22}+\ro_{33}-2\re \ro_{23})\\
&\ro_{34}(t)=-e^{-2\g
 t}(\ro_{12}+\ro_{13})+\frac{1}{2}e^{-\g
 t}(2\ro_{12}+2\ro_{13}+\ro_{24}+\ro_{34})-\frac{1}{2}(\ro_{24}-
\ro_{34})\\
&\ro_{44}(t)=-\frac{1}{2}e^{-2 \g t}(1+\ro_{11}-\ro_{44}+2\re
\ro_{23})-2\g t e^{-2\g
t}\ro_{11}+\frac{1}{2}(1+\ro_{11}+\ro_{44}+2\re \ro_{23})
\end{split}
\end{equation*}
 and remaining matrix elements can be obtained by hermiticity
condition $\ro_{kj}=\overline{\ro}_{jk}$. In the limit $t \to
\infty$ we obtain asymptotic (stationary) states parametrized as
follows \begin{equation} \ro_{\mathrm{as}}=\begin{pmatrix}
0&\hspace{2mm}0&\hspace{2mm}0&\hspace{2mm}0\\
0&\hspace{2mm}\al&-\al&\hspace{2mm}\be\\
0&-\al&\hspace{2mm}\al&-\be\\
0&\hspace{2mm}\overline{\be}&-\overline{\be}&1-2\al
\end{pmatrix}
\end{equation}
where
\begin{equation}
\al=\frac{1}{4}(\ro_{22}+\ro_{33}-2\re \ro_{23}),\quad
\be=\frac{1}{2}(\ro_{24}-\ro_{34})
\end{equation}
We can also compute concurrence of the asymptotic state and
 the result is:\\[2mm]
\textit{
Concurrence  of asymptotic state of the semi
- group $\{ T_{t} \}$ generated by $L$ with $g=1$ equals to
\begin{equation}
C(\ro_{\mathrm{as}})=2|\al|=\frac{1}{2}|\ro_{22}+\ro_{33}-2\re \ro_{23}|
\end{equation}
where $\ro_{jk}$ are the matrix elements of the initial
state.}\\[2mm]
\section{Some examples}
\noindent In this section we consider examples of initial states
and its evolution.\\[2mm]
\textbf{I. Pure separable states}\\[2mm]
Let
\begin{equation}
\ro=P_{\Psi\otimes \Phi}=P_{\Psi}\otimes P_{\Phi}
\end{equation}
where
$$
\Psi=\begin{pmatrix}
\Psi_{1}\\
\Psi_{2}
\end{pmatrix}
\in \cH_{A},\quad
\Phi=\begin{pmatrix}
\Phi_{1}\\
\Phi_{2}
\end{pmatrix}\in \cH_{B}
$$
are normalized. Then one can check that
\begin{equation}
\begin{split}
\al&=\frac{1}{4}(1-|\ip{\Psi}{\Phi}|^{2})\\
\be&=\frac{1}{2}(|\Phi_{2}|^{2}\Psi_{1}\overline{\Psi}_{2}-|\Psi_{2}|^{2}
\Phi_{1}\overline{\Phi}_{2})
\end{split}
\end{equation}
where $\ip{\cdot}{\cdot}$ is the inner product in $\C^{2}$. So
\begin{equation}
C(\ro_{\mathrm{as}})=\frac{1}{2}(1-|\ip{\Psi}{\Phi}|^{2})
\end{equation}
From the formula (21) we see that there are separable initial
states for which asymptotic states are entangled. In particular,
the asymptotic state has a maximal concurrence if vectors $\Psi$
and $\Phi$ are orthogonal and its concurrence is zero (the state
remains separable) if $|\ip{\Psi}{\Phi}|=1$.
\par
\noindent
Now we discuss some special cases.\\
\textbf{a.}When one atom is in excited state and the other is in 
ground
state
$$
\Psi=\ket{1},\quad \Phi=\ket{0}
$$
the asymptotic (mixed) state is given by
$$
\ro_{\mathrm{as}}=\begin{pmatrix}
0&\hspace{2mm}0&\hspace{2mm}0&0\\[2mm]
0&\hspace{2mm}\frac{1}{4}&-\frac{1}{4}&0\\[2mm]
0&-\frac{1}{4}&\hspace{2mm}\frac{1}{4}&0\\[2mm]
0&0&0&0
\end{pmatrix}
$$
It can also be shown that in this case the relaxation process
produces the state $\ro_{t}$ with concurrence
$$
C(\ro_{t})=\frac{1-e^{-\g t}}{2}
$$
increasing to the maximal value equal to $1/2$. Thus two atoms
initially in separable state become entangled for all $t$ and the
asymptotic (steady) state attains the maximal amount of
entanglement.\\
 \textbf{b.} When two atoms are in excited states
$$
\Psi=\Phi=\ket{1}
$$
the asymptotic state equals to
$$
\ket{0}\otimes \ket{0}
$$
Thus the relaxation process brings two atoms into ground
states.\\
\textbf{c.} The state $\ket{0}\otimes\ket{0}$ is stationary state
for semi - group $\{ T_{t} \}$.\\[4mm]
\textbf{II. Pure maximally entangled states}\\[2mm]
Let
\vskip 2mm
\noindent
$$
\ro=Q(a,\theta_{1},\theta_{2})=\begin{pmatrix}
\frac{a^{2}}{2}&
\frac{a\sqrt{1-a^{2}}}{2}e^{-i\theta_{1}}&
\frac{a\sqrt{1-a^{2}}}{2}e^{-i\theta_{2}}&
-\frac{a^{2}}{2}e^{-i(\theta_{1}+\theta_{2})}\\[2mm]
\frac{a\sqrt{1-a^{2}}}{2}e^{i\theta_{1}}&\frac{1-a^{2}}{2}&
\frac{1-a^{2}}{2}e^{i(\theta_{1}-\theta_{2})}&-\frac{a\sqrt{1-a^{2}}}{2}e^{-
i\theta_{2}}\\[2mm]
\frac{a\sqrt{1-a^{2}}}{2}e^{i\theta_{2}}&\frac{1-a^{2}}{2}e^{-i(\theta_{1}-
\theta_{2})}&
\frac{1-a^{2}}{2}&-\frac{a\sqrt{1-a^{2}}}{2}e^{-i\theta_{1}}\\[2mm]
-\frac{a^{2}}{2}e^{i(\theta_{1}+\theta_{2})}&-\frac{\sqrt{1-
a^{2}}}{2}e^{i\theta_{2}}&
-\frac{a\sqrt{1-a^{2}}}{2}e^{i\theta_{1}}&\frac{a^{2}}{2}
\end{pmatrix}
$$
\vskip 2mm
\noindent
where $ a\in [0,1],\; \theta_{1},\theta_{2}\in
[0,2\pi]$. Pure states $Q(a,\theta_{1},\theta_{2})$ are maximally
entangled and form a family of all maximally entangled states of
the $4$ - level system \cite{BJO1}. It turns out that
$\ro_{\mathrm{as}}$ is defined by
\begin{equation}
\begin{split}
\al&=\frac{1}{4}(1-a^{2})(1-\cos (\theta_{1}-\theta_{2}))\\
\be&=\frac{1}{4}a\sqrt{1-a^{2}}(e^{-i\theta_{1}}-e^{-i\theta_{2}})
\end{split}
\end{equation} and \begin{equation} 
C(\ro_{\mathrm{as}})=\frac{1}{2}(1-a^{2})(1-\cos
(\theta_{1}-\theta_{2})) \end{equation} From the formula (23) we see 
that
there are initial maximally entangled states which asymptotically
become separable ($a=1$ or $\theta_{1}-\theta_{2}=2k\pi$) and
such that the asymptotic concurrence is greater then $0$. States
with $a=0$ and $\theta_{1}-\theta_{2}=(2k+1)\pi$ remain maximally
entangled. For example the state \begin{equation}
\frac{1}{\sqrt{2}}(\ket{0}\otimes\ket{1}-\ket{1}\otimes\ket{0})
\end{equation} is stable. On the other hand, the concurrence of 
\begin{equation}
\frac{1}{\sqrt{2}}(\ket{0}\otimes\ket{1}+\ket{1}\otimes\ket{0})
\end{equation} goes to zero faster then the concurrence of 
\begin{equation}
\frac{1}{\sqrt{2}}(\ket{0}\otimes\ket{0}+\ket{1}\otimes\ket{1})
\end{equation} as shown on \textbf{Fig. 1.} below. In Dicke's theory 
of
spontaneous radiation processes the state (24) is called
subradiant whereas the state (25) has half the lifetime of a
single atom and therefore is called superradiant \cite{Dicke}. We
see that the time-dependence of concurrence  reflects the
relaxation properties of those states. \vskip 8mm \noindent
\begin{picture}(300,300)
\put(50,70){\begin{picture}(180,180) \epsffile{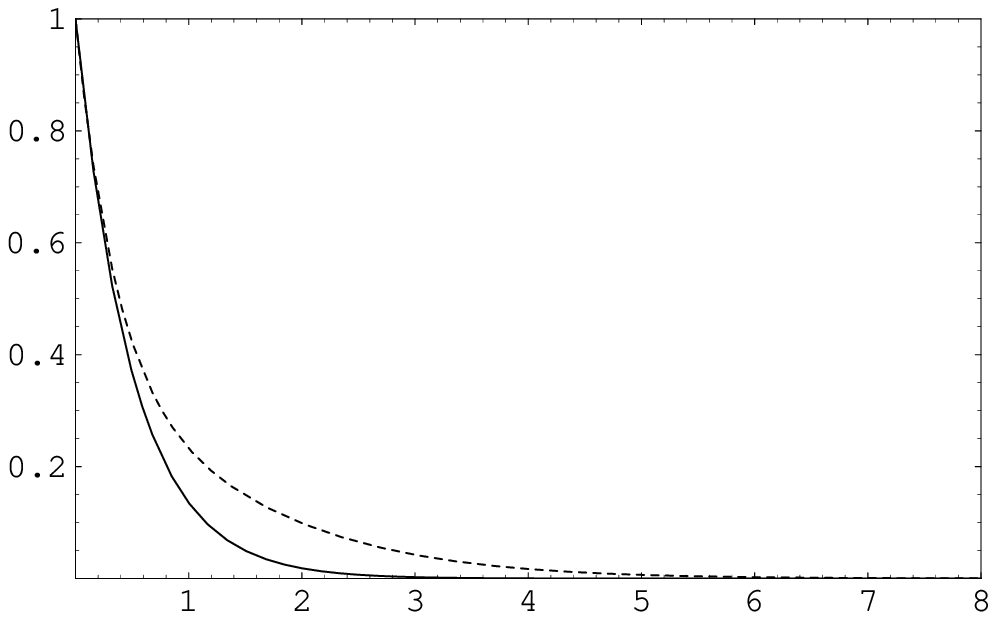}
\end{picture}}
\put(350,75){$\g t$} \put(60,255){$C(\ro_{t})$}
\end{picture}
\vskip -18mm \noindent \centerline{\textbf{Fig. 1.} Concurrence as
the function of time for initial states:
$\frac{1}{\sqrt{2}}(\ket{0}\otimes\ket{0}+\ket{1}\otimes\ket{1})$}
\hspace*{20mm} (dotted line)
and
$\frac{1}{\sqrt{2}}(\ket{0}\otimes\ket{1}+\ket{1}\otimes\ket{0})$
(solid line).
\vskip 6mm
\noindent
\textbf{III. Some classes of mixed states}\\[2mm]
\textbf{a.} Bell - diagonal states. Let \begin{equation}
\ro_{\mathrm{B}}=p_{1}\ket{\Phi^{+}}\bra{\Phi^{+}}+p_{2}\ket{\Phi^{-
}}\bra{\Phi^{-}}+p_{3}
\ket{\Psi^{+}}\bra{\Psi^{+}}+p_{4}\ket{\Psi^{-}}\bra{\Psi^{-}}
\end{equation} where Bell states $\Phi^{\pm}$ and $\Psi^{\pm}$ are 
given by
\begin{equation} 
\Phi^{\pm}=\frac{1}{\sqrt{2}}(\ket{0}\otimes\ket{0}\,\pm\,
\ket{1}\otimes\ket{1}),\quad
\Psi^{\pm}=\frac{1}{\sqrt{2}}(\ket{1}\otimes\ket{0}\,\pm\,
\ket{0}\otimes\ket{1}) \end{equation} It is known that all $p_{i}\in
[0,1/2],\; \ro_{\mathrm{B}}$ is separable, while for
$p_{1}>1/2,\; \ro_{\mathrm{B}}$ is entangled with concurrence
equal to $2p_{1}-1$ (similarly for $p_{2},\,p_{3}\, p_{4}$)
\cite{HHHa}. Now the asymptotic state has the form 
\begin{equation}
\ro_{\mathrm{as}}=\begin{pmatrix}
0&\hspace{2mm}0&\hspace{2mm}0&0\\[2mm]
0&\hspace{2mm}\frac{p_{4}}{2}&-\frac{p_{4}}{2}&0\\[2mm]
0&-\frac{p_{4}}{2}&\hspace{2mm}\frac{p_{4}}{2}&0\\[2mm]
0&\hspace{2mm}0&\hspace{2mm}0&1-p_{4}
\end{pmatrix}
\end{equation}
with  concurrence $C(\ro_{\mathrm{as}})=p_{4}$. So even when the
initial state is separable, the asymptotic state becomes
entangled. \\[2mm]
\textbf{b.} Werner states \cite{BBPSSW}. Let
\begin{equation}
\ro_{\mathrm{W}}=(1-p)\frac{\I_{4}}{4}+p\ket{\Phi^{+}}\bra{\Phi^{+}}
\end{equation}
If $p>1/3,\; \ro_{\mathrm{W}}$ is entangled with concurrence equal
to $(3p-1)/2$. On the other hand
\begin{equation}
\ro_{\mathrm{as}}=\begin{pmatrix}
0&0&0&0\\[2mm]
0&\frac{1-p}{8}&\frac{p-1}{8}&0\\[2mm]
0&\frac{p-1}{8}&\frac{1-p}{8}&0\\[2mm]
0&0&0&\frac{3+p}{4}\\[2mm]
\end{pmatrix}
\end{equation}
has the concurrence $C(\ro_{\mathrm{as}})=\frac{1-p}{4}$, so the
asymptotic states are entangled for all $p\neq 1$. Notice that
even completely mixed state $\frac{\I_{4}}{4}$ evolves to
entangled asymptotic state.\\[2mm]
\textbf{c.} Maximally entangled mixed states . The
 states
\begin{equation}
\ro_{\mathrm{M}}=\begin{pmatrix}
h(\delta)&0&0&\delta/2\\[2mm]
0&1-2h(\delta)&0&0\\[2mm]
0&0&0&0\\[2mm]
\delta/2&0&0&h(\delta)
\end{pmatrix},\quad h(\delta)=\begin{cases}
1/3& \delta\in [0,2/3]\\
\delta/2& \delta\in [2/3,1]
\end{cases}
\end{equation}
are conjectured to be maximally entangled for a given degree
of inpurity measured by $\tr \ro^{2}$ \cite{Kwiat}. According to
(18) the concurrence of the asymptotic state is given by
\begin{equation}
C(\ro_{\mathrm{as}})=\frac{1}{2}(1-2h(\delta))
\end{equation}
\vskip 8mm
\noindent
\begin{picture}(300,300)
\put(50,70){\begin{picture}(180,180) \epsffile{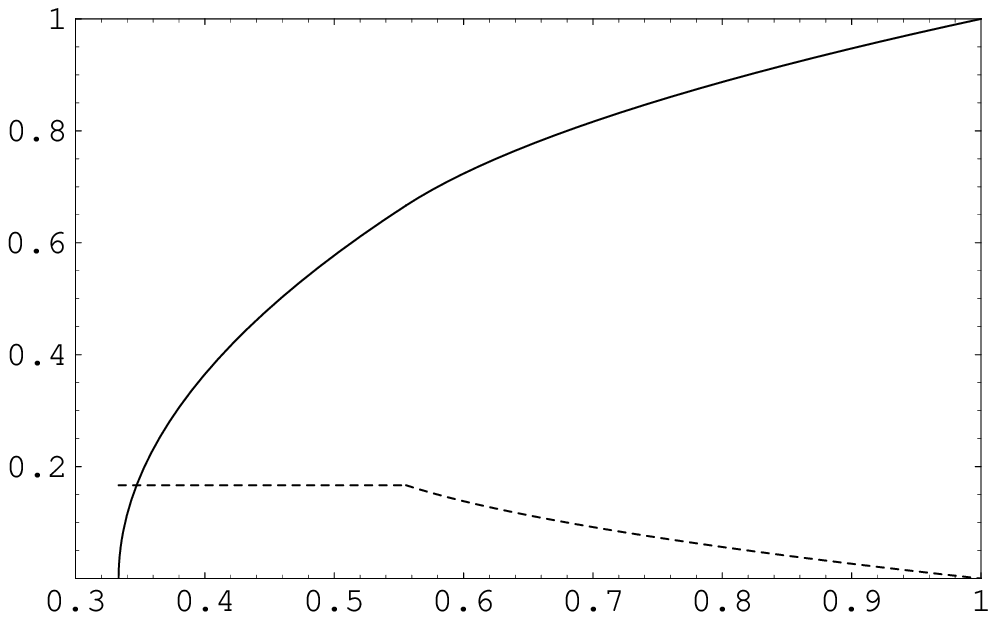}
\end{picture}}
\put(350,75){$\tr \ro_{\mathrm{M}}^{2}$}
\put(60,255){$C(\ro_{t})$}
\end{picture}
\vskip -18mm \noindent \centerline{\textbf{Fig. 2.} Concurrence of
$\ro_{\mathrm{M}}$ (solid line) and $\ro_{\mathrm{as}}$ (dotted
line) as the function of $\tr \ro_{\mathrm{M}}^{2}$} \vskip 6mm
\noindent Even in that case, there are initial states (for
sufficiently small $\tr \ro_{\mathrm{M}}^{2}$) such that the
asymptotic state is more entangled (see \textbf{Fig. 2.}).
\section{Remarks on general case}
In the case of arbitrary distance between the atoms i.e. when
$g\in [0,1)$, semi-group generated by $L$ is uniquely relaxing,
with the asymptotic state $\ket{0}\otimes\ket{0}$. Thus, for any
initial state $\ro$, the concurrence $C(\ro_{t})$ approaches $0$
when $t\to \infty$. But it does not mean that the function $t\to
C(\ro_{t})$ is always monotonic. The general form of $C(\ro_{t})$
is rather involved, so we consider only some special cases.\\[2mm]
\textbf{1.} Let the initial state of the compound system be equal
to $\ket{0}\otimes\ket{1}$. This states evolves to \vskip 4mm
\noindent
\begin{equation}
\ro_{t}=\begin{pmatrix} 0&0&0&0\\[2mm]
0&\frac{1}{2}e^{-\g t}\,(\cosh \ga t +1)&-\frac{1}{2}e^{-\g
t}\sinh \ga t&0\\[2mm]
0&-\frac{1}{2}e^{-\g t}\sinh \ga t& \frac{1}{2}e^{-\g t}\, (\cosh
\ga t -1)&0\\[2mm]
0&0&0&1-e^{-\g t}\cosh \ga t
\end{pmatrix}
\end{equation} \vskip 4mm \noindent with concurrence 
\begin{equation} C(\ro_{t})=e^{-\g
t}\sinh \ga t \end{equation} In the interval $[0,t_{\ga}]$, where
$$
t_{\gamma}=\frac{1}{2\ga}\ln \frac{\g +\ga}{\g -\ga}
$$
the function (34) is increasing to its maximal value
$$
C_{\mathrm{max}}=\frac{\ga}{\g-\ga}\,
\left(\frac{\g+\ga}{\g-\ga}\right)^{-\frac{\DS\g+\ga}{\DS 2\ga}}
$$
 whereas for
$t>t_{\ga}$,  $C(\ro_{t})$ decreases to $0$. Thus for any nonzero
photon exchange rate $\ga$, dynamics given by the semi - group
$\{ T_{t} \}$ produces some amount of entanglement between two
atoms which are initially in the ground state and excited state.
Note that the maximal value of $C(\ro_{t})$ depends only on
emission rates $\g$ and $\ga$.\\[2mm]
\textbf{2.} For the initial states
$$
\Psi^{\pm}=\frac{1}{\sqrt{2}}\,(\ket{0}\otimes\ket{1}\pm
\ket{1}\otimes\ket{0})
$$
the relaxation to the asymptotic state $\ket{0}\otimes\ket{0}$ is
given by density matrices \vskip 4mm\noindent
\begin{equation}
\ro_{t}^{\pm}=\begin{pmatrix} 0&0&0&0\\[2mm]
0&\frac{1}{2}e^{-(\g \pm \ga)t}&-\frac{1}{2}e^{-(\g\pm
\ga)t}&0\\[2mm]
0&-\frac{1}{2}e^{-(\g\pm\ga)t)}&\frac{1}{2}e^{-(\g\pm\ga)t}&0\\[2mm]
0&0&0&1-e^{-(\g\pm\ga)t}
\end{pmatrix}
\end{equation}
\vskip 4mm\noindent with the corresponding concurrence
$$
C(\ro_{t}^{\pm})=e^{-(\g\pm\ga)t}
$$
The state $\Psi^{-}$ is no longer stable (as in the case of
$\ga=\g$), but during the evolution its concurrence goes to zero
slower than $C(\ro^{+}_{t})$ (\textbf{Fig. 3.}). For $\ga $ close
to $\g$, $\Psi^{-}$ is almost stable.
\vskip 8mm
\noindent
\begin{picture}(300,300)
\put(50,70){\begin{picture}(180,180) \epsffile{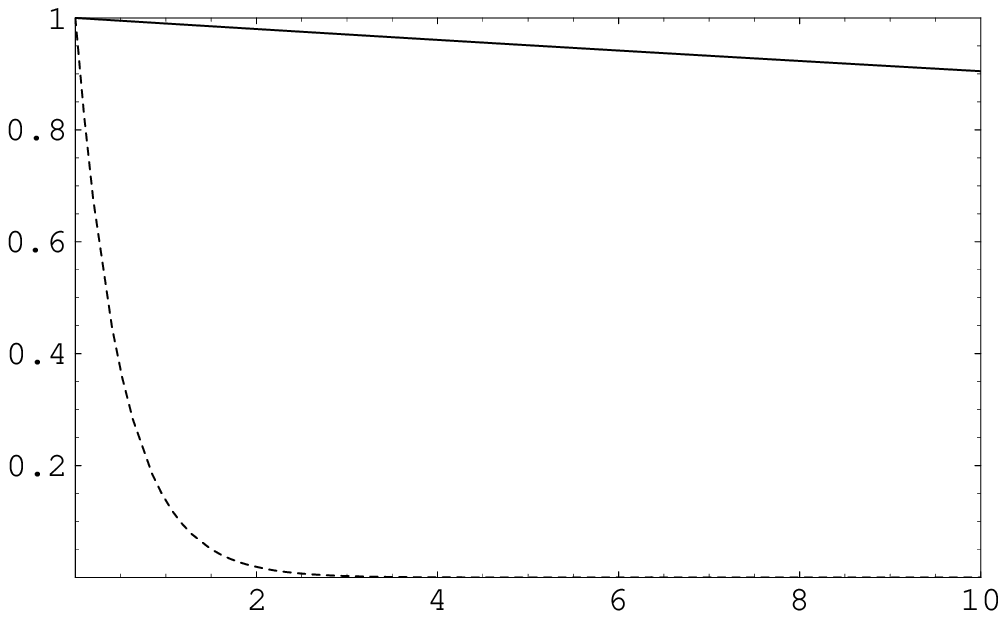}
\end{picture}}
\put(350,75){$\g t$} \put(60,255){$C(\ro_{t})$}
\end{picture}
\vskip -18mm
\noindent
\centerline{\textbf{Fig. 3.}
$C(\ro^{+}_{t})$ (dotted line) and $C(\ro^{-}_{t})$ (solid line)
for $\ga/\g = 0.99$} 
\end{document}